\documentclass[aps,twocolumn,showpacs,amsmath,amssymb]{revtex4}
\usepackage{graphicx} 
\usepackage{dcolumn} 
\usepackage{bm} 

\begin{document}


\title{Thin Film Formation During Splashing of Viscous Liquids}

\author{Michelle M.\ Driscoll} 
\author{Cacey S.\ Stevens}

\author{Sidney R.\ Nagel} 
\affiliation{ The James Franck Institute and Department of Physics, The University of Chicago }%

\date{\today}

\begin{abstract} After impact onto a smooth dry surface, a drop of viscous liquid initially spreads in the form of a thick lamella.  If the drop splashes, it first emits a thin fluid sheet that can ultimately break up into droplets causing the splash.   Ambient gas is crucial for creating this thin sheet.  The time for sheet ejection, $t_{ejt}$, depends on impact velocity, liquid viscosity, gas pressure and molecular weight.  A central air bubble is trapped below the drop at pressures even below that necessary for this sheet formation.  In addition, air bubbles are entrained underneath the spreading lamella when the ejected sheet is present.  Air entrainment ceases at a lamella velocity that is independent of drop impact velocity as well as ambient gas pressure.
\end{abstract}


\maketitle

\section{\label{sec:level1}Introduction}

When a liquid drop hits a dry surface, there are many possible outcomes; it can rebound completely as one drop, spread smoothly, or splash dramatically ejecting many smaller droplets.   Although this appears superficially similar to the impact of a drop on a thin liquid layer, that process is quite different; there much of the ejected crown originates from the surface layer \cite{joss_zales_thinfilm,Siggi_ejetasht,yarinWeiss_thinfilm,howison,deegan}.  As one might expect, upon hitting a dry surface liquid viscosity, drop size, impact velocity, surface tension, and substrate elasticity and roughness play an important role in creating a splash. Previous work has defined a boundary between the splashing and spreading states based solely on these criteria \cite{Mundo, Range, Stow, Pepper,YarinWeiss}.  This characterization of a splashing threshold implicitly assumes that this list of control parameters is not only complete but also that the physical mechanism for splashing is essentially the same in all cases studied.  Thus, one might expect that the splash threshold should scale monotonically with those parameters. However, recent experiments call both these assumptions into question.  First, there is an additional crucial control parameter for creating a splash.  Lowering the ambient gas pressure below a threshold value can completely suppress splashing:  above the threshold pressure a splash is seen, while below it, the drop spreads out smoothly, without breaking apart into secondary droplets \cite{LeiPRL}.  This indicates that gas pressure is key to the mechanism of splash creation.  Second, the threshold pressure for splashing depends non-monotonically on the viscosity of the liquid \cite{LeiPRE}.  The threshold pressure first decreases and then turns around and begins increasing as the viscosity is increased.  This non-monotonic scaling indicates that as viscosity increases, different stabilizing (or destabilizing) forces come into play.  These two regimes in viscosity therefore must be examined separately.  

\begin{figure}[htbp]
\centering 
\begin{center} 
\includegraphics[width=3.4in]{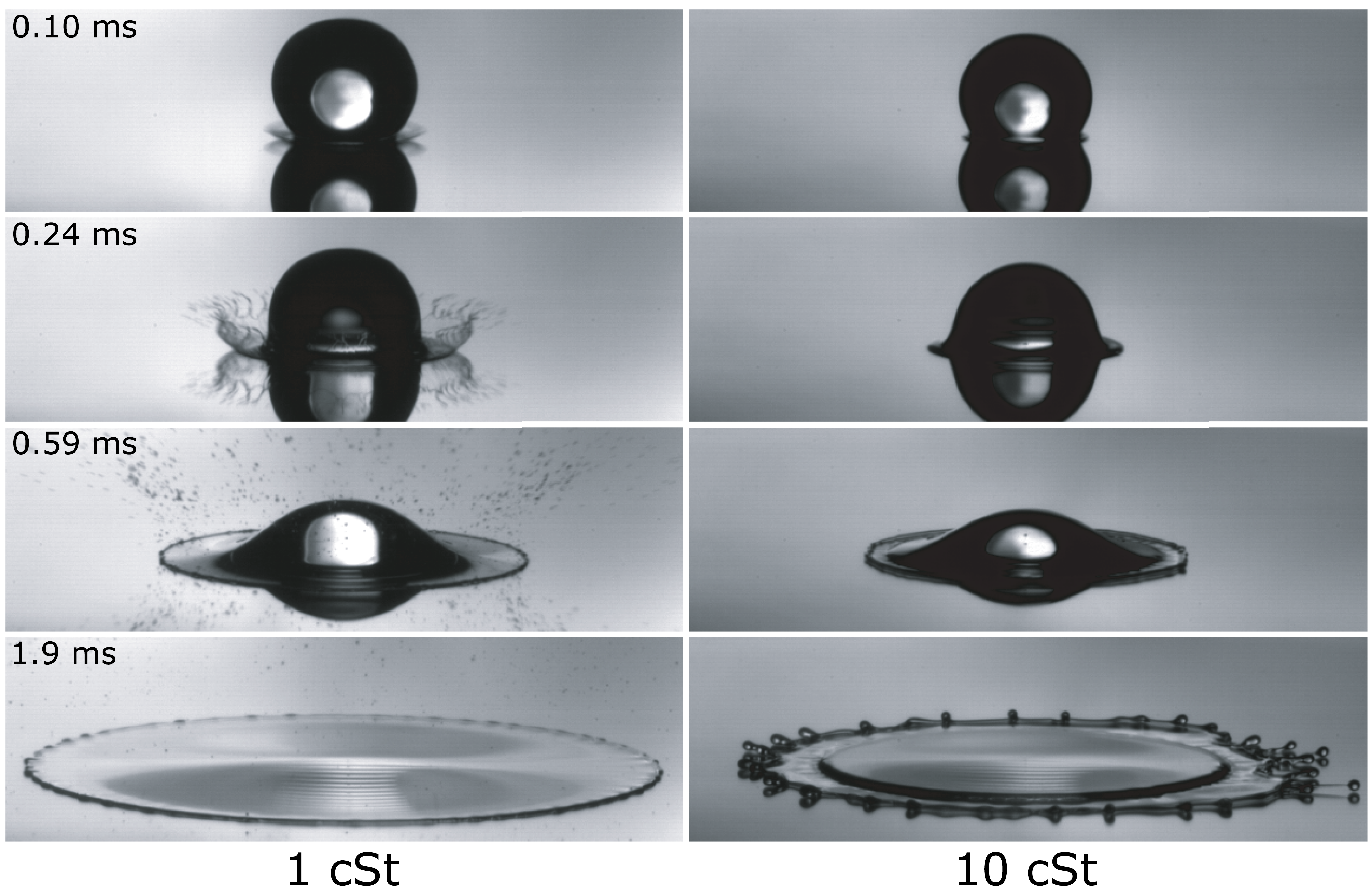} 
\end{center}
\caption{Splash in the low- and high-viscosity regimes.  The left (right) column shows the impact of a 3.4 mm (3.1 mm) $1$ cSt (10 cSt) drop of silicone oil hitting a smooth, dry microscope slide.  Both drops impact at $3.55 \pm0.04$ m/s.  There is a striking difference in the two splashing behaviors.  Splashing occurs much later in the viscous case, and the liquid sheet is ejected at a much shallower angle.  The sheet ejection time, $t_{ejt}$, occurs between the 2nd and 3rd panel in the right column. } 
\label{lowVShigh}
\end{figure}

We will show here that splashing when the liquid viscosity, $\nu_L$, is large exhibits substantially different behavior than when $\nu_L$ is low.  For high viscosities, splashing is delayed with respect to low viscosity liquids, as shown in Figure \ref{lowVShigh}.  In the low $\nu_L$ regime, experimentally determined scaling relations support a model where the compressibility of the surrounding gas becomes important for creating a splash \cite{LeiPRL}.  It is assumed that compressibility effects are important at times very close to impact, when the lamella is expanding rapidly.  However, viscous splashing only starts to be observed at late times, when the lamella velocity $\sim$ 3 m/s.   Compressibility effects are highly unlikely to be relevant at a Mach number of 0.01 (in air).

After impact, a viscous drop spreads out smoothly for several tenths of a millisecond, giving no indication of an impending splash. Suddenly, at a time $t_{ejt}$, a very thin sheet of liquid is ejected from the spreading lamella, as seen in the third image in the right column of Figure \ref{lowVShigh}.  This thin sheet travels outward nearly parallel to the substrate.  This is strikingly different from the low $\nu_L$ case, where the corona lifts off the substrate at a large angle  \cite{ZamoraDFD08}. Furthermore, in the high $\nu_L$ regime, the thin liquid sheet can be ejected and persist without breaking up into secondary droplets.   We note here the splashing observed for silicone oil at low viscosity  (shown in the left column of Figure \ref{lowVShigh}) is similar in all regards to what has been previously reported in the splashing of ethanol \cite{LeiPRL}.  This corroborates  that the volatility of the liquid is unimportant \cite{XuPRL_reply}.

Thin sheet formation is  the precursor to splashing in viscous liquids.  By looking solely at droplet emission \cite{Mundo, Range, Stow, Pepper,YarinWeiss}, only the breakup of the thin sheet is examined, not its creation.  Here, we report on a detailed study of the sheet formation.  We measure how $t_{ejt}$ depends on the various control parameters.  We also report on gas entrainment at the lamellar edge \cite{DriscollDFD09,Siggi} and how it is linked to the appearance of the thin sheet.  

\section{Experimental Details}

For $3.1$ mm diameter drops, the onset of the high $\nu_L$ splashing regime occurs at $\nu_L$ $\sim$  3 cSt \cite{LeiPRE}. We used two kinds of viscous liquids in our studies.  Silicone oils (PDMS, Clearco Products) allow the liquid viscosity, $\nu_L$, to be varied between $3$ cSt and $50$ cSt while keeping the surface tension, $\sigma$, nearly constant between $19.7 - 20.8$ dyn/cm.  Mixtures of  water and  glycerin can also be made over the same viscosity range but with higher surface tension $\sigma = 67$ dyn/cm.  Using a syringe pump (Razel Scientific, Model R99-E), pumping at $10.2$ - $15.9$ cc/hour we generated drops of reproducible diameter, $d_0$ = $3.1 \pm 0.1$ mm.  The drops were released from rest in a vacuum chamber whose pressure could be varied between $1$ kPa and $101$ kPa.  By varying the release height from $0.1$ m to $6$ m, we controlled the impact velocity on the substrate, $u_0$, from $1.1$ m/s to $8.7$ m/s.  This varies the kinetic energy, $E_k = \frac{1}{2} \rho \left[ \frac{4}{3} \pi  \left( \frac{d_0}{2} \right)^3 \right] u_0^2$, in the drop from 11.21 - 552.0 $\mu$J.  For the 10 cSt silicone oil, the Reynolds number, $\mbox{Re}$, ($\mbox{Re}$ = inertial forces / viscous forces) varied from 380 to 2700.  The Weber number, $\mbox{We}$, ($\mbox{We}$ = inertial forces / surface tension forces) varied from 220 - 11,000.  The substrates were dry, smooth, glass surfaces (Fisherbrand Cover Glass).  A new surface was used for each drop to ensure that there was no contamination from previous trials.  We filmed the impacts at speeds up to $97,000$ fps using a Phantom v12, Vision Research camera.  Video images were used to determine droplet emission, sheet formation time, $t_{ejt}$, drop diameter, $d_0$, and impact velocity, $u_0$.  

\section{Threshold Pressure}

\begin{figure}[htbp]
\centering
\begin{center} 
\includegraphics[width=3.4in]{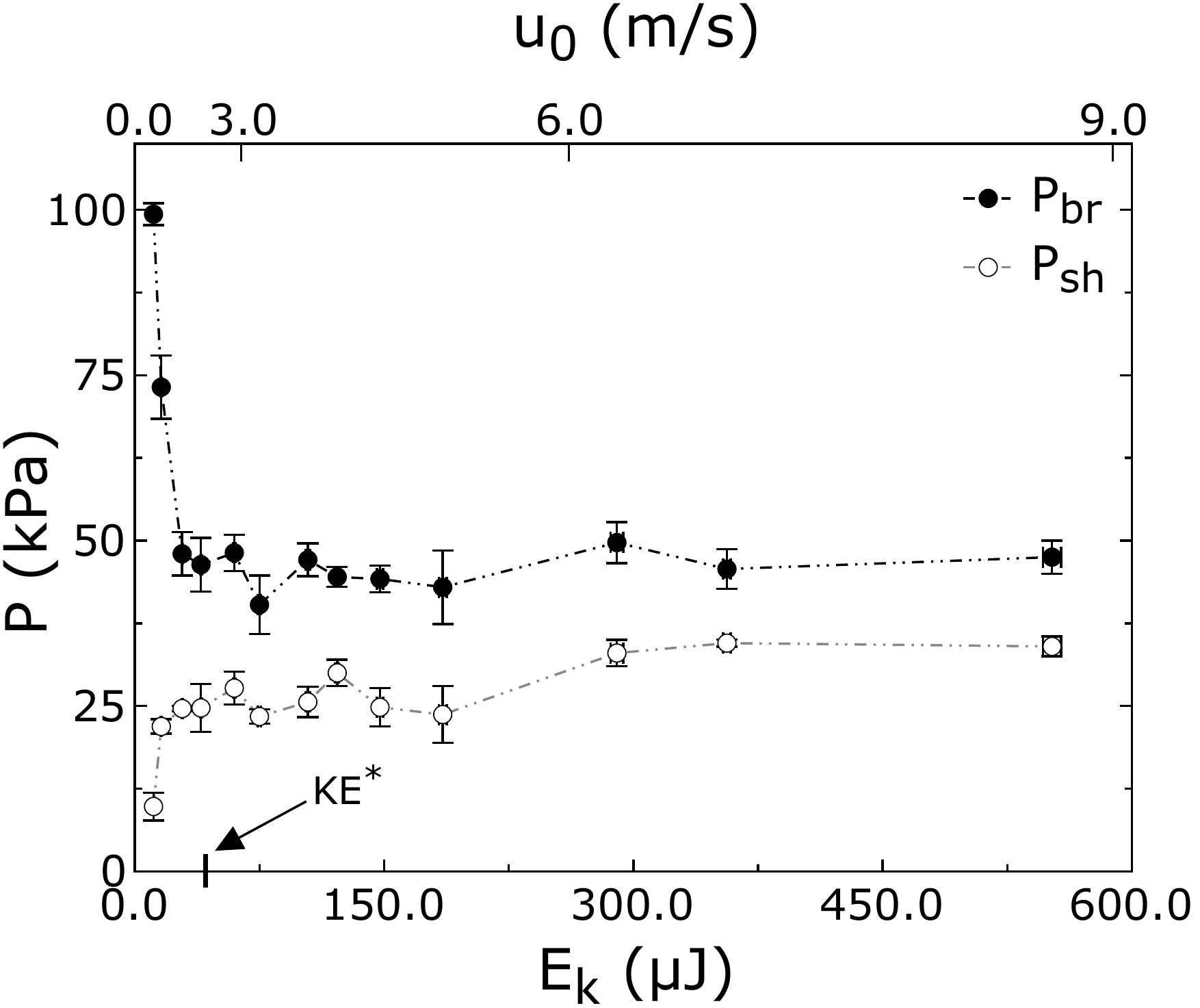} 
\end{center}
\caption{Threshold pressures $P_{sh}$, $P_{br}$ versus kinetic energy (lower axis) and impact velocity (upper axis) for 10 cSt silicone oil.  $P_{br}$, ($\circ$), the threshold pressure for breakup into secondary droplets is insensitive to impact velocity / kinetic energy above $E_k^* \sim$  30 $\mu$J / $u^* \sim$ 2 m/s.  $P_{sh}$, ($\bullet$), the threshold pressure for thin sheet formation slowly increases above $E_k^*$ / $u^*$.  The error bars for $P_{sh}$ ($P_{br}$) are defined by the range of pressures over which sheet ejection (droplet breakup) first begins to occur.  Dashed lines are a guide to the eye.} 
\label{P_{br}}
\end{figure}

One control parameter for the ejection of a thin sheet is the ambient gas pressure.  Figure \ref{P_{br}} shows that at low pressure, the drop spreads in a lamella but never emits a thin sheet.  This regime has been observed in simulations \cite{Schroll}.  When the pressure is raised, a thin sheet is formed at $P_{sh}$ as shown by the lower curve in Figure \ref{P_{br}}.  While undulations may appear on the rim of this thin sheet \cite{LeiPRE}, no secondary droplets are formed and no splashing occurs.  When the pressure is raised above $P_{br}$, the sheet breaks into secondary droplets; the boundary for droplet break-off is shown by the upper curve.  When droplets are emitted, they are formed both on the thickened rim of the sheet and also from the sheet ripping apart, giving rise to a large distribution of droplet sizes.  Below $P_{br}$, no droplets are emitted.  $P_{sh}$ is determined by the pressure at which the ejection of a thin sheet is first detected, and $P_{br}$ is determined by the pressure at which droplets break off.  The determination of both $P_{sh}$ and $P_{br}$  can vary over a range as broad as 10 kPa between drops with ostensibly the same control parameters.  This is shown by the error bars in Figure \ref{P_{br}}:  the lower (upper) bound is set by the lowest (highest) pressure at which sheet or drop ejection is observed.

\begin{figure}[htbp] 
\centering
\begin{center} 
\includegraphics[width=3.4in]{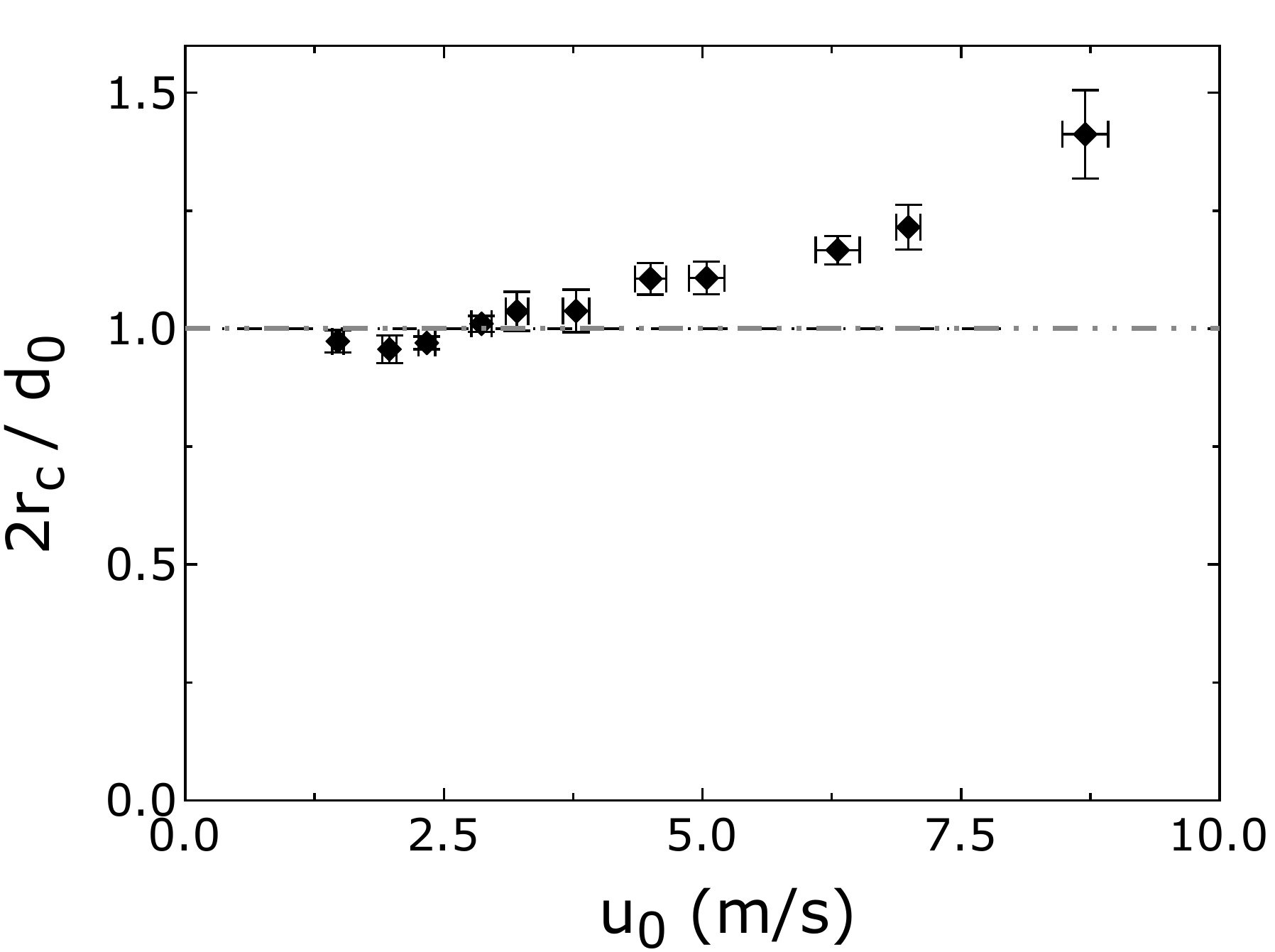} 
\end{center}
\caption{The variation in drop shape with velocity at 30 kPa (just above $P_{sh}$).  Air drag has an increasing effect on the shape of the falling drop as velocity is increased.  Plotted is $2r_c$/$d_0$ versus $E_k$, where 2$r_c$ is twice the radius of curvature of the bottom edge of the drop,  and we have normalized it by the drop diameter, $d_0$.  The dashed line indicates 2$r_c$ = $d_0$. } 
\label{dropshape}
\end{figure}

Figure \ref{P_{br}} shows two ostensibly different behaviors, separated by a characteristic kinetic energy, $E_k^*$ $\sim$ 30 $\mu$J or velocity, $u^* \sim$ 2.3 m/s.  We define $E_k^*$ / $u^*$ as the threshold above which $P_{br}$ is essentially flat. The behavior of $P_{br}$ and $P_{sh}$ both seen counterintuitive;  $E_k$ increases by more than an order of magnitude, but $P_{br}$ remains the same and $P_{sh}$ increases slightly.  Over this range of $u_0$, the drop shape varies considerably due to air drag.  Figure 3 shows that the radius of curvature at the bottom of the impacting drop, $r_c$, increases with $u_0$.  This larger effective radius may affect the dependence of $P_{sh}$ on $u_0$.

\begin{figure}[htbp] 
\centering
\begin{center} 
\includegraphics[width=3.4in]{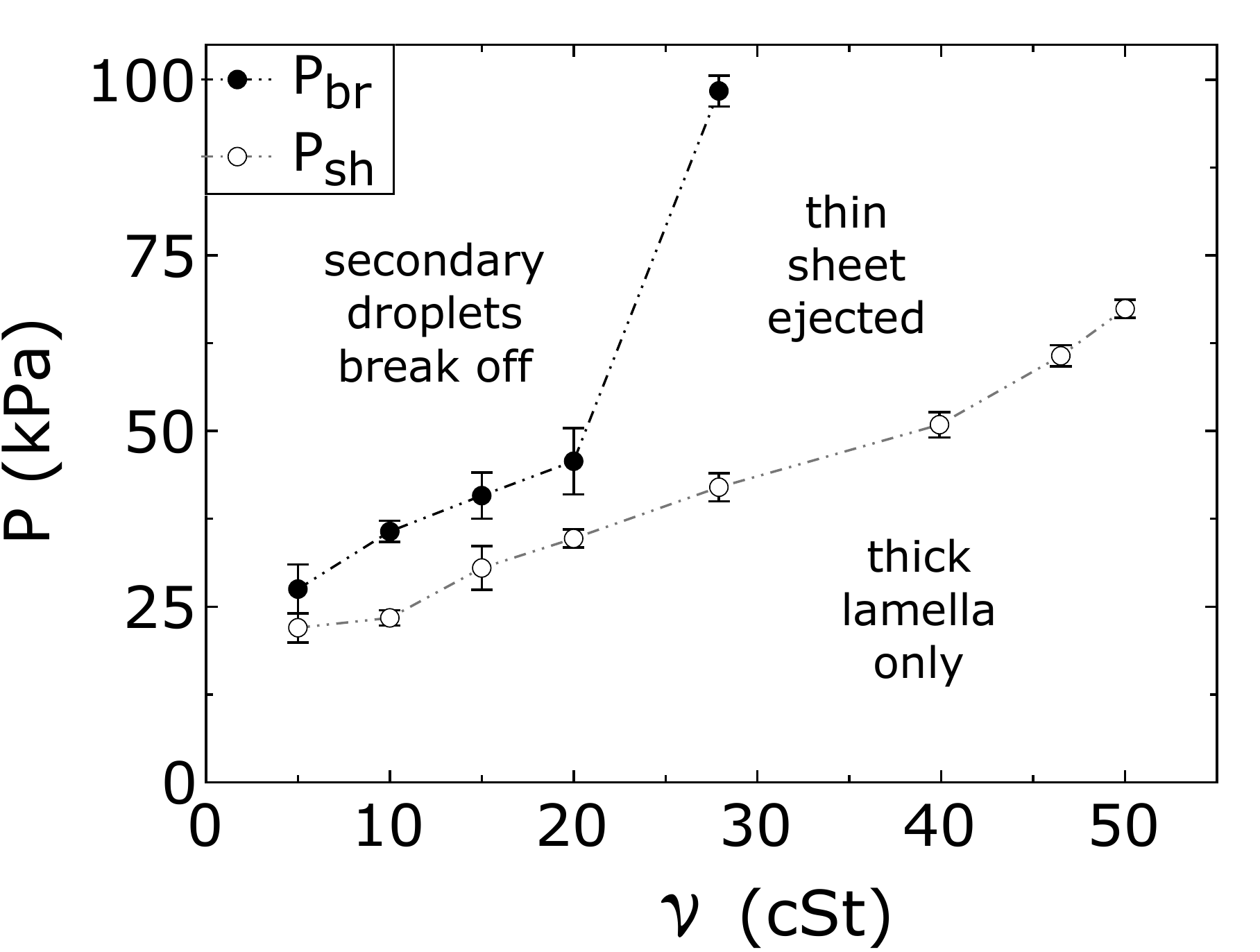} 
\end{center}
\caption{The threshold pressures $P_{br}$ ($\bullet$) and $P_{sh}$ ($\circ$) versus viscosity. As viscosity increases, both $P_{br}$ and $P_{sh}$ increase monotonically. The three separate regions of splashing behavior are labeled. As viscosity increases there is a widening region between $P_{br}$ and $P_{sh}$.  Above $\nu$ $\sim$ 30 cSt, only the spreading and sheet-only states are present; no droplets are ever emitted in in this regime.   Dashed lines are a guide to the eye.} 
\label{PtvsNu}
\end{figure}

\begin{figure}[htbp] 
\centering
\begin{center} 
\includegraphics[width=3.4in]{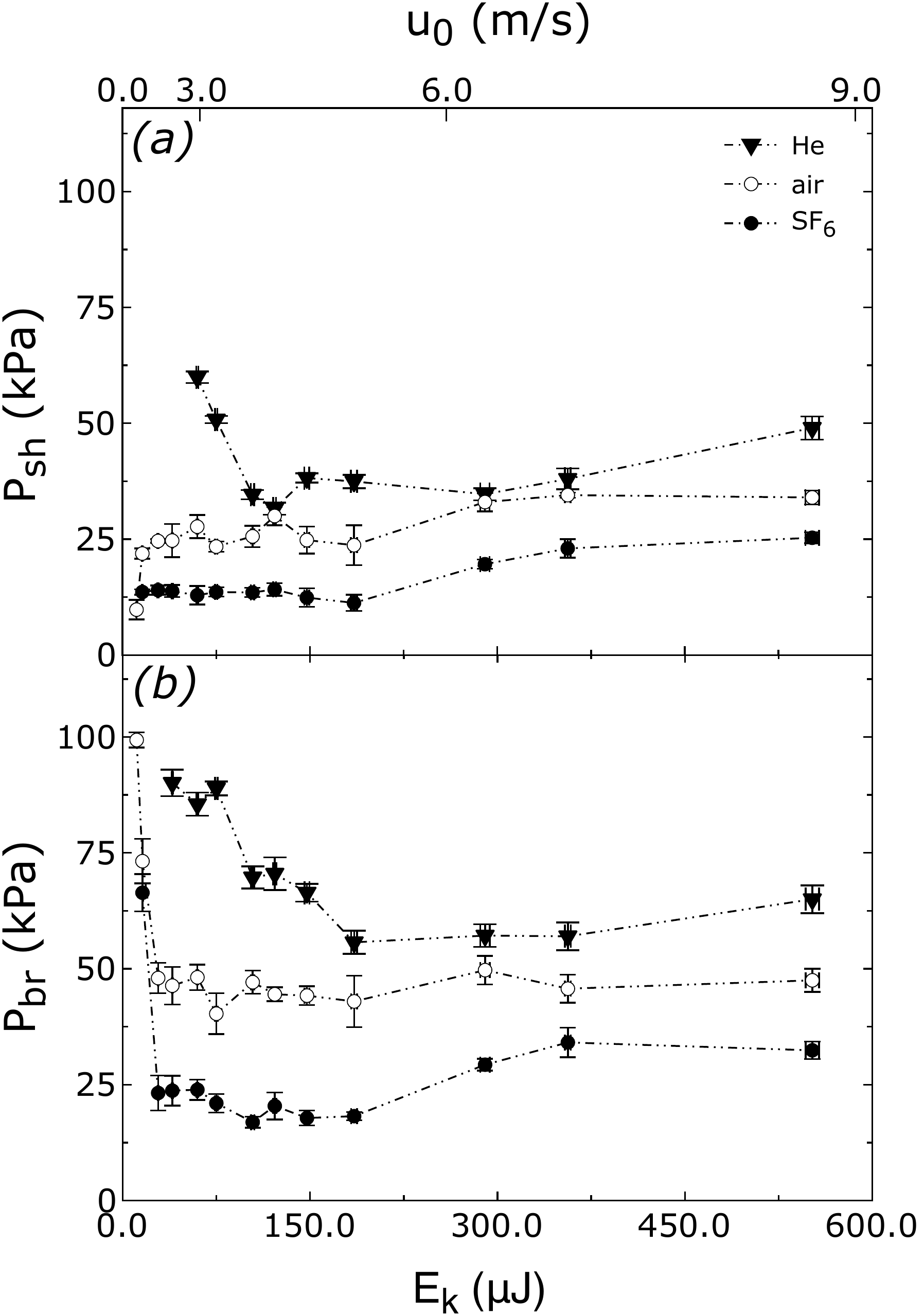} 
\end{center}
\caption{$(a)$ $P_{sh}$ and $(b)$ $P_{br}$ versus $E_k$ and $u_0$ for gases of different molecular weight, $M_G$ for 10 cSt silicone oil drops with $d_0$ = 3.1 $\pm$ 0.1 mm.  Three gases were used: He ($\blacktriangledown$), $M_G$ = 4 Da, air ($\circ$), $M_G$ = 29 Da, and SF$_6$ ($\bullet$), $M_G$ = 146 Da.  The behavior for air and SF$_6$ appear similar for both $P_{sh}$ and $P_{br}$, except the threshold pressures are lower in SF$_6$.  For both gases, $u^* \sim$ 2.0 m/s.  In He, $P_{br}$ only becomes relatively insensitive to $E_k$ above 125 $\mu$J ($u_0 \sim $4 m/s).  Dashed lines are a guide to the eye.}
\label{PtGas} 
\end{figure}

To further examine how thin sheet emission can occur without splashing, we varied $\nu_L$ while holding other parameters constant by using silicone oils of various molecular weights. (We kept $u_0 = 3.21 \pm 0.03$ m/s and $d_0 = 3.1 \pm 0.1$ mm.  For the oils used, $\sigma_L$ varies only slightly from 19.7 - 20.8 dyn/cm.)  Xu has previously shown evidence of the dependence of threshold pressure on viscosity \cite{LeiPRE}, but did not examine the sheet regime separately. Xu defined a pressure, $P_{bump}$, below which undulations no longer appear on the rim of the spreading lamella.  We too see these undulations but they appear on the thin sheet only after it has expanded.  The appearance of the thin sheet is distinctly different from the onset of undulations.

As shown in Figure \ref{PtvsNu}, both $P_{br}$ and $ P_{sh}$ increase monotonically with increasing viscosity between $\nu_L$ = 5 and 50 cSt.  As viscosity increases, the two threshold pressures move further apart, producing a widening region where the sheet-only state occurs.  

The ambient gas pressure, $P$, determines whether or not a thin sheet will form.  To see if $P$ is the only property of the gas that acts as a control parameter, we studied three gases with different molecular weights, $M_G$: He ($M_G$ = 4 Da), air ($M_G$ = 29 Da), and SF$_6$ ($M_G$ = 146 Da), while keeping $\nu_L$ = 10 cSt and $d_0 = 3.1 \pm 0.1$ mm.  Figure \ref{PtGas}$a$ shows $P_{sh}$ and Figure \ref{PtGas}$b$ shows $P_{br}$ versus $E_k$ and $u_0$ for the three gases.  Above $E_k$ = 15 $\mu$J, $P_{sh}$ behaves the same in both air and SF$_6$ in that it is nearly independent of kinetic energy.  $P_{sh}$ in He appears to have similar behavior, but only above $E_k$ = 125 $\mu$J.  Below that, $P_{sh}$ in He increases rapidly with lowering $E_k$.  This behavior is qualitatively different than the other gases: $P_{sh}$ in air decreases rapidly with lowering $E_k$, while in He it increases.  $P_{br}$ also appears very similar in air and SF$_6$, and has a crossover behavior near $E_k$ = 30 $\mu$J.  At low $E_k$, $P_{br}$ in He decreases much less rapidly than in air or SF$_6$, but above $E_k$ = 185 $\mu$J, it displays similar behavior to the other two gases.

\section{Thin Sheet Formation}

\begin{figure}[htbp]
\centering 
\begin{center} 
\includegraphics[width=3.4in]{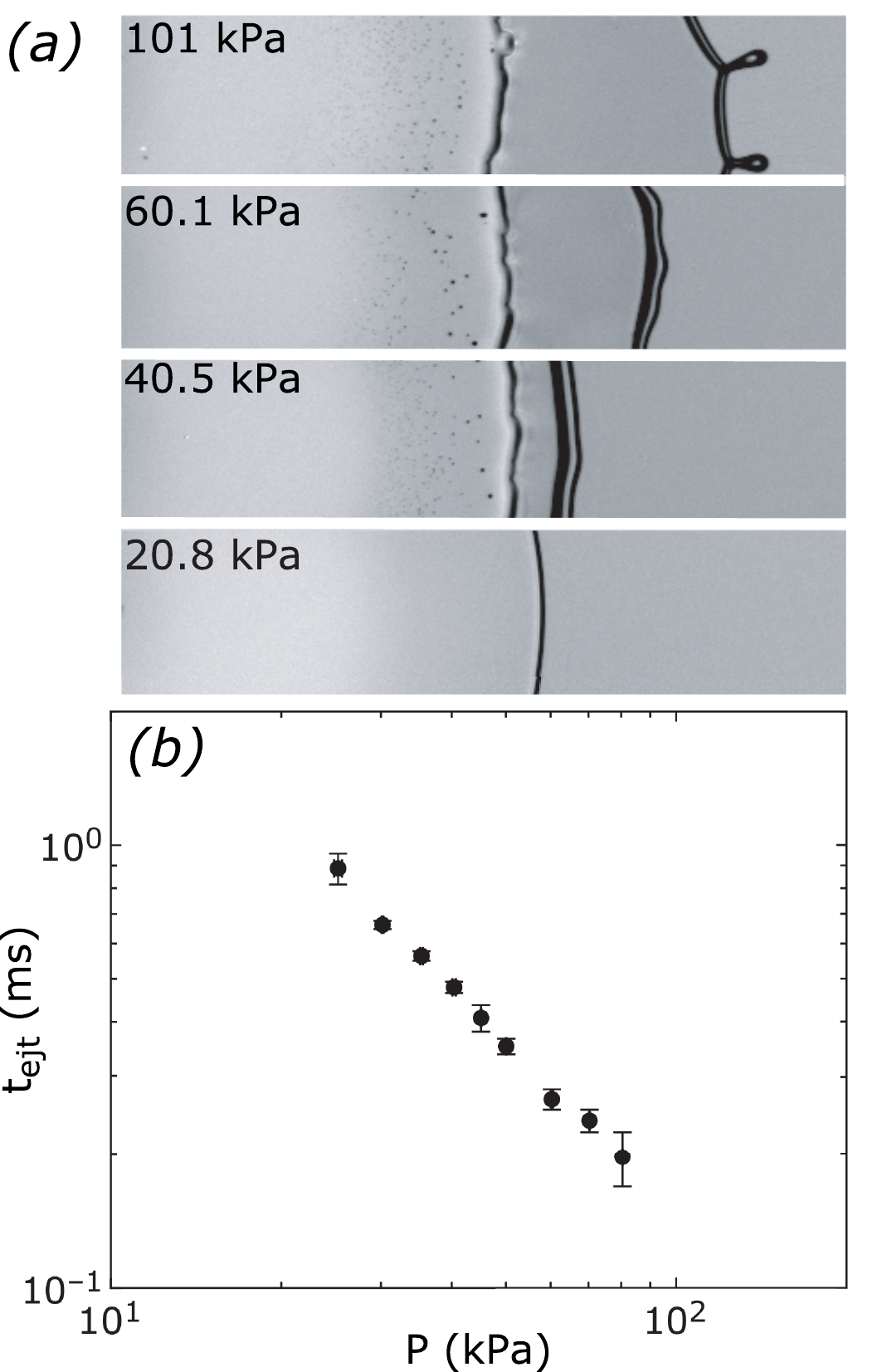} 
\end{center}
\caption{Variation of thin sheet ejection with pressure.  $(a)$  The images are photographs taken from below of a drop of silicone oil ($d_0$ = 3.1 mm, $E_k$ = 59.8 $\mu$J ($u_0$ = 2.86 m/s), $\nu$ = 10 cSt) at the same time after impact (1.55 ms). Pressure is labeled in the upper left corner of each image.  The variation in the extent of the thin sheet is due to its appearance at a later time (longer $t_{ejt}$).  $(b)$  Thin sheet ejection time versus pressure.  As the ambient pressure is lowered, thin sheet ejection time becomes delayed.}
\label{tejtvsP} 
\end{figure}

The images in Figure \ref{tejtvsP}$a$ show how the thin sheet evolves as a function of pressure.  As the ambient pressure is decreased, the ejection time, $t_{ejt}$, increases until no sheet ejection occurs below $P_{sh}$.  This is shown quantitatively in Figure \ref{tejtvsP}$b$.  This behavior contrasts with that in the low $\nu_L$ regime where, within the temporal resolution of experiments, the corona (if present) always appears at the moment of impact \cite{LeiPRL} (also see first image in Figure \ref{lowVShigh}).  Although lowering the gas pressure does supress splash formation in this low $\nu_L$ regime, it does so by creating a smaller corona, not by ejecting it at a later time.

To determine what other parameters, aside from the gas pressure, set the time scale $t_{ejt}$, we systematically varied several control parameters.  Impact velocity, $u_0$, was varied from $1.97 - 6.20$ m/s.  Gas pressure, $P$, was varied from $1$ kPa to $101$ kPa.  Gas molecular weight, $M_G$, was varied by using three different gases: He, air, and SF$_6$.  Liquid viscosity, $\nu_L$, was varied from 5 cSt to 50 cSt.  Liquid surface tension was varied by using a water and glycerin solution ($\sigma$ = 67 dyn/cm) in addition to various silicone oils ($\sigma$ = 19.7 - 20.8 dyn/cm).  

\begin{figure}[htbp] 
\begin{center} 
\includegraphics[width=3.4in]{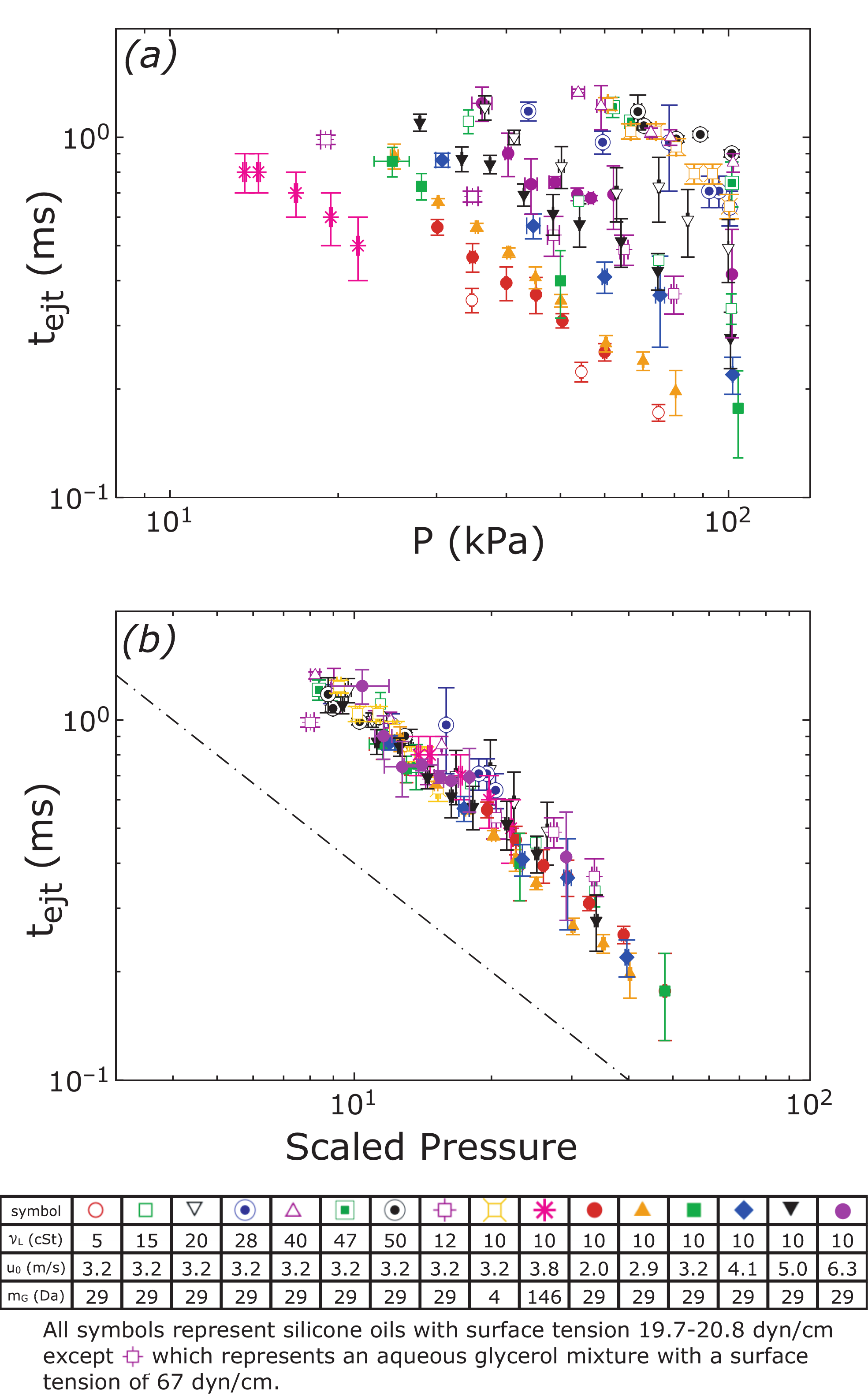} 
\end{center}
\caption{$(a)$  Sheet ejection time versus pressure, varying $u_0$, $\sigma$,  $\nu_{L}$, and $M_G$.  Symbols corresponding to the data sets are given in the table.  $(b)$ Data collapse using a scaled pressure, $P_{scaled} = m_{G}^{0.56} \nu_{L}^{-0.8} u_0^{-0.7} P$ ($m_G$ in Da, $\nu_L$ in cSt, and $u_0$ in m/s).  The scaling was found empirically as described in the text.  The dashed line indicates a power law with exponent $-1$.} 
\label{tvsP} 
\end{figure}

As shown in Figure \ref{tvsP}$a$, all of the individual data sets follow $t_{ejt} \sim P^{-.94 \pm .22}$. Varying the parameters  $M_G$, $u_0$, $\sigma$, and $\nu_{L}$ only shifts the $\log{t_{ejt}}$ versus $\log{P}$ curve, suggesting that the data can be collapsed onto a master curve.  To determine the scaling factors for data collapse, each parameter set was treated individually.  For example, data sets of varying viscosity were separately fit to $t_{ejt} = aP^{-1}$, generating a set of prefactors, $\{ a \}_{\nu_L}$.  These prefactors were then fit to $a_{\nu_L} \sim \nu_L^n$, and this exponent $n$ was used to collapse the data onto the master curve.  This procedure was then repeated for the remaining parameter sets: $M_G$, $u_0$, and $\sigma$.  Using this procedure we find $t_{ejt}$ scales as $\nu_L^{-0.8 \pm 0.1}$, $u_0^{-0.7 \pm 0.1}$, and $M_G^{0.56 \pm 0.05}$ (where $M_G$ is in Da, $\nu_L$ is in cSt, and $u_0$ is in m/s).  The error bars indicate the confidence interval for the fit in each parameter set.  Each parameter set has a small dynamic range, only about a decade in time and pressure.  The data appear to be insensitive to surface tension within error, and therefore no power law could be reliably fit to this parameter.  The data collapse is then given by $P_{scaled} = M_G^{0.56} \nu_{L}^{-0.8} u_0^{-0.7} P$.

\section{Bubble Entrapment at Impact}

\begin{figure}[htbp] 
\centering 
\begin{center} 
\includegraphics[width=3.2in]{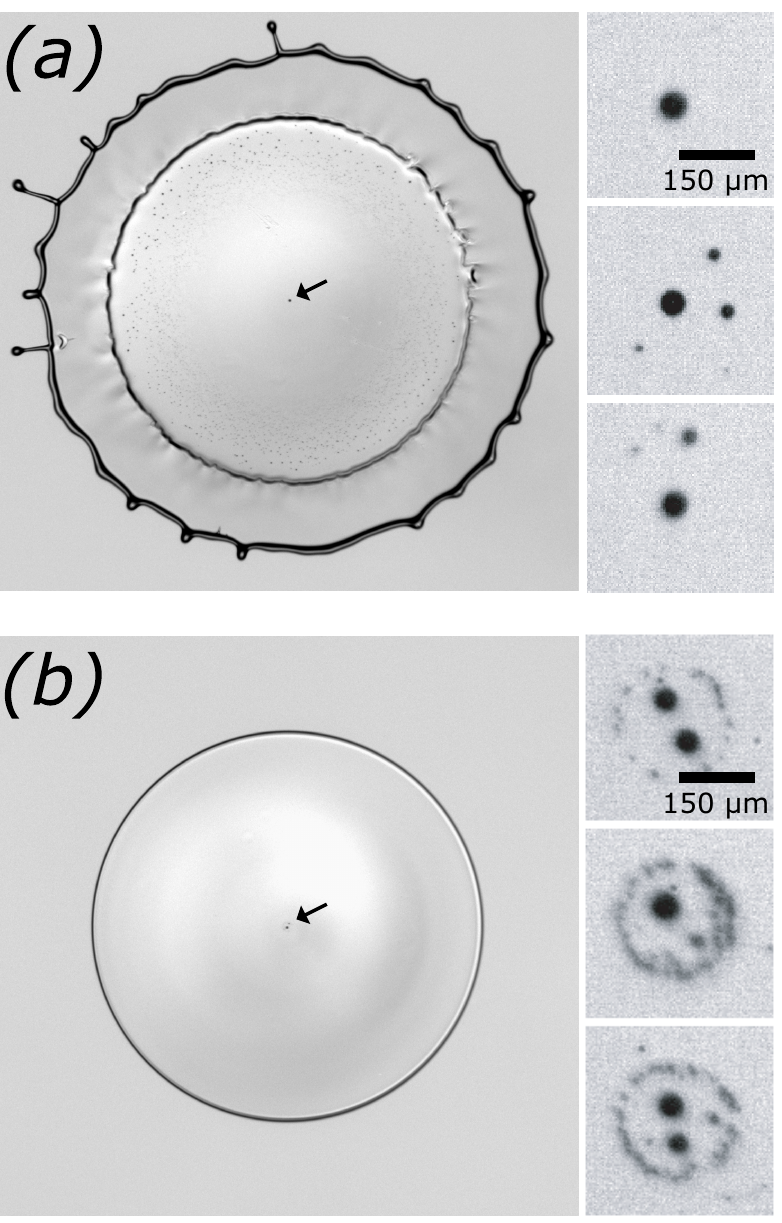} 
\end{center}
\caption{Air entrapment during impact.  The images show silicone oil drops $\sim$ 2 ms after impact ($\nu_L$ = 10 cSt, $u_0$ = 3.19 m/s, $d_0$ = 3.1 mm). The large image provides an overall view, with the arrow indicating the central bubble. The three smaller images are magnified views of this central region. The three smaller images represent three different impacts under nominally the same conditions. The smaller images are slightly blurred due to their large magnification.  \textit{(a)}  $Atmospheric$ $pressure$:  The entrapped air forms one large central bubble, and often several distinct smaller bubbles.  \textit{(b)}  $P$ $\sim$ $20$ $kPa$:  The central bubble is still present and  a ring of micro-bubbles is also entrapped.  The transition to creating this ring of micro-bubbles is not abrupt, but occurs over a wide range in pressure.} 
\label{ctrbubble} 
\end{figure}

When a drop impacts a surface, a small amount of gas is trapped underneath that then contracts into a bubble.  This is a robust phenomenon and has been observed in a variety of systems, including both high and low $\nu_L$ liquids \cite{Siggi_fingering,Siggi_center,Chandra2,vanDam}.  The entrapment of a bubble  underneath the impacting drop has been observed in simulations as well, and has been suggested as a mechanism for splash formation in low $\nu_L$ liquids \cite{Mandre,BrennerJFM,Hicks}.  In our studies, we observe that the entrapped bubble is not adhered to the substrate, but slowly rises up into the liquid, popping when it reaches the top surface of the lamella.  This popping occurs long after the drop has finished spreading, $\sim$ 20 ms after impact.  Here we report on the effect of air pressure on the central entrapped bubble.  

Figure \ref{ctrbubble} shows two drops of 10 cSt silicone oil $\sim$ 2 ms after impact.  Figure \ref{ctrbubble}$a$ shows the center bubble entrapped at atmospheric pressure, while Figure \ref{ctrbubble}$b$ shows the bubble entrapped at $\sim$ 20 kPa, which is below $P_{sh}$.  A central bubble is always observed, even below $P_{sh}$, when the drop does not create a thin sheet or emit droplets.  At lower pressures  a ring of micro-bubbles is entrapped in addition to one or more larger bubbles.  The entrapment of a central bubble versus a central bubble plus micro-bubble ring does not appear to have a threshold pressure.  Rather, as pressure is lowered there is a gradual transition: entrapment of a micro-bubble ring occurs more frequently and the entrapped micro-bubbles are larger at lower pressures.  Such rings of micro-bubbles have been observed in both the low and high viscosity regimes \cite{Siggi_center}.  

Although the ambient pressure changes the morphology of the entrapped air pocket, a central bubble of gas is nevertheless entrapped in all three impact outcomes: droplet break-off plus thin sheet ejection, sheet ejection only, and lamella spreading only.  The formation of an ejected thin sheet occurs long after the entrapped bubble has collapsed, and furthermore, the shape of the entrapped bubble has no strong variation as the pressure is varied across $P_{sh}$.  It is therefore not clear whether or how air entrapment upon impact is connected to the thin sheet formation in the high viscosity regime.

\section{Bubble Entrainment During Spreading}

\begin{figure}[htbp] 
\centering 
\begin{center} 
\includegraphics[width=2.8in]{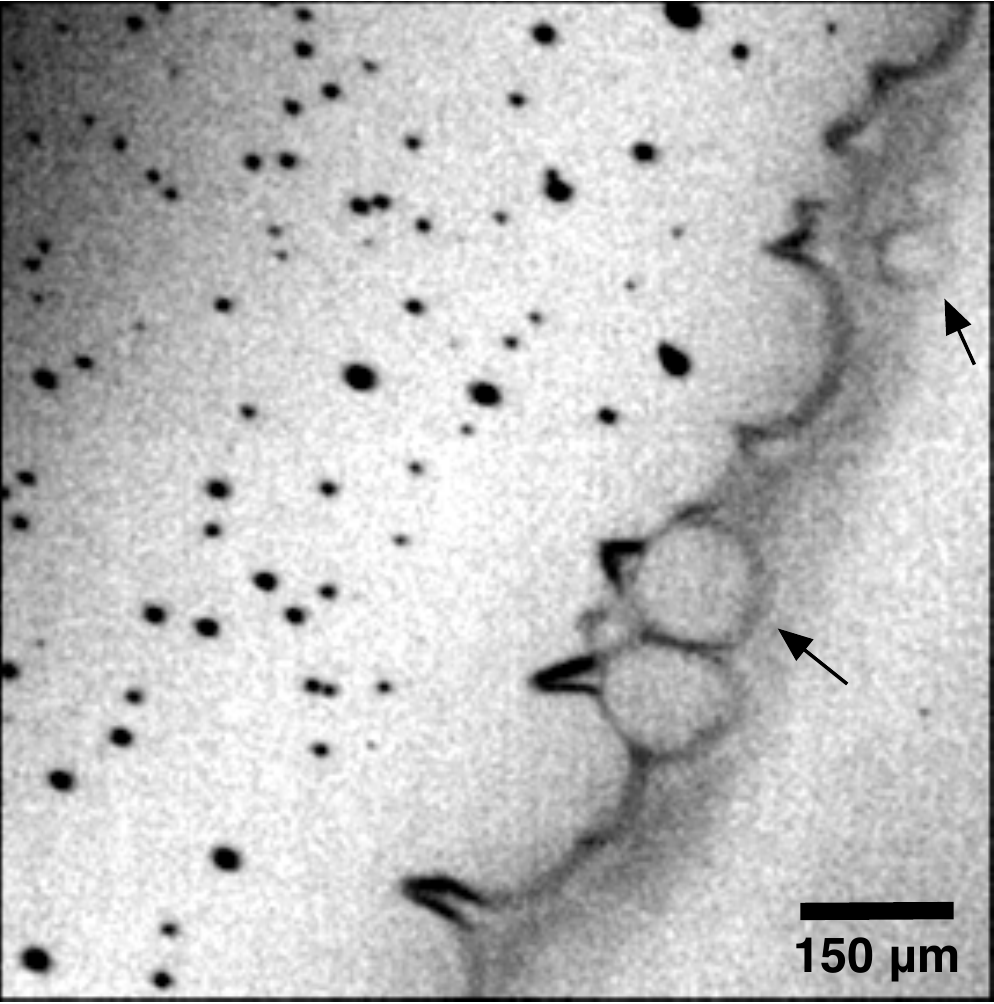} 
\end{center}
\caption{A magnified view of the lamella / thin sheet interface, after a thin sheet has formed during spreading  (silicone oil drop: $\nu_L$ = 10 cSt, $u_0$ = 3.19 m/s, $d_0$ = 3.1 mm). Gas bubbles are entrained at the lamella/thin sheet interface when the thin sheet locally makes contact with the substrate, as shown by the arrows.  The upper arrow indicates a thin-sheet contact event that just occurred.  The lower arrow indicates a pair of events that have trapped a pocket of air behind them, thus creating a bubble.  Bubble entrainment only occurs when the thin sheet is present; below $P_{sh}$ the lamella edge remains smooth and no gas bubbles are entrained (see Figure \ref{ctrbubble}$b$).} 
\label{5xbubble} 
\end{figure}

As reported previously and shown in Figure \ref{5xbubble}, small gas bubbles are entrained underneath the edge of the spreading lamella \cite{DriscollDFD09,Siggi}.  We show here that these bubbles are only present after the thin sheet is emitted.  In order to check that these bubbles were due to entrainment of gas during spreading, and not dissolved gas coming out of solution, all liquids were thoroughly degassed before the impact experiments were done.   Furthermore, the same syringe of liquid is seen to exhibit bubble entrainment above but not below $P_{sh}$.

Bubble entrainment begins when the thin sheet emerges, at the time $t_{ejt}$.  As shown in Figure \ref{5xbubble}, air entrainment occurs through local contact where the thin sheet touches down to meet the substrate.  These contacts form small circular regions where the thin sheet wets the substrate, as indicated by the arrows in Figure \ref{5xbubble}.  These regions always appear at or very near the lamella/thin sheet interface, and they appear continuously from the time the thin sheet is ejected, $t_{ejt}$, until bubble entrainment ceases at $t_{stop}$.  Bubble entrainment occurs as pockets of air are trapped behind these regions.  The increase in bubble size with time is due to the fact that the individual area of these local contact regions increases in time. This resembles one of the mechanisms proposed by Thoroddsen et. al \cite{Siggi}, who associated this with imperfections in the smoothness of the substrate.  In contrast, we associate this entrapment with the presence of the thin sheet, which helps focus the air to the lamella / substrate interface.

\begin{figure}[htbp] 
\centering 
\begin{center} 
\includegraphics[width=3.4in]{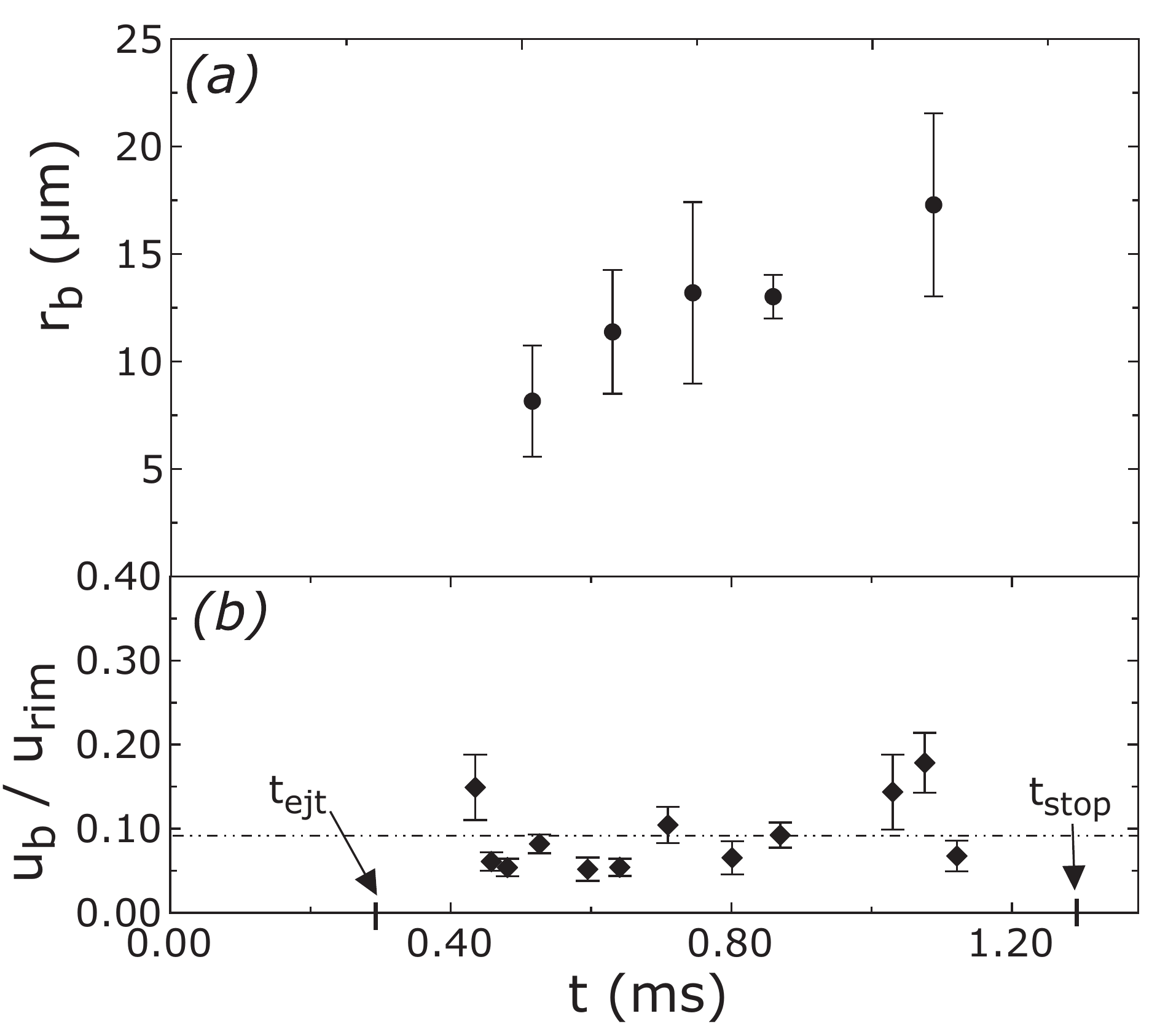} 
\end{center}
\caption{$(a)$  Average bubble radius, $r_b$, at time of entrainment versus time after impact, $t$.  As the lamella is spreading, larger and larger bubbles are entrained.  $(b)$ Velocity of the entrained bubbles, $u_b$, normalized by the local lamella velocity, $u_{rim}$, at the time of entrainment. The dashed line indicates the average of $u_b$/$u_{rim}$.  The entrained bubbles are not adhered to the substrate, but have an initial radial velocity $\sim$ 10\% that of the lamella edge.} 
\label{bubb_RandV} 
\end{figure}

Larger and larger bubbles are entrained until a time $t_{stop}$, at which time gas entrainment suddenly ceases.  The time $t_{stop}$ occurs before the thin sheet is ripped off of the lamella.  Figure \ref{bubb_RandV}$a$ shows the average bubble radius, $r_b$, vs. time.  At $t_{stop}$, the average bubble radius is $\sim$ 17 $\mu$m.  The entrained bubbles are not adhered to the substrate below them --- they move outward, but at an initial velocity, $u_b$ which is $\sim$ 10\%  that of the lamella velocity, $u_{rim}$, regardless of bubble size, as shown in Figure \ref{bubb_RandV}$b$.

\begin{figure}[htbp] 
\centering 
\begin{center} 
\includegraphics[width=3.4in]{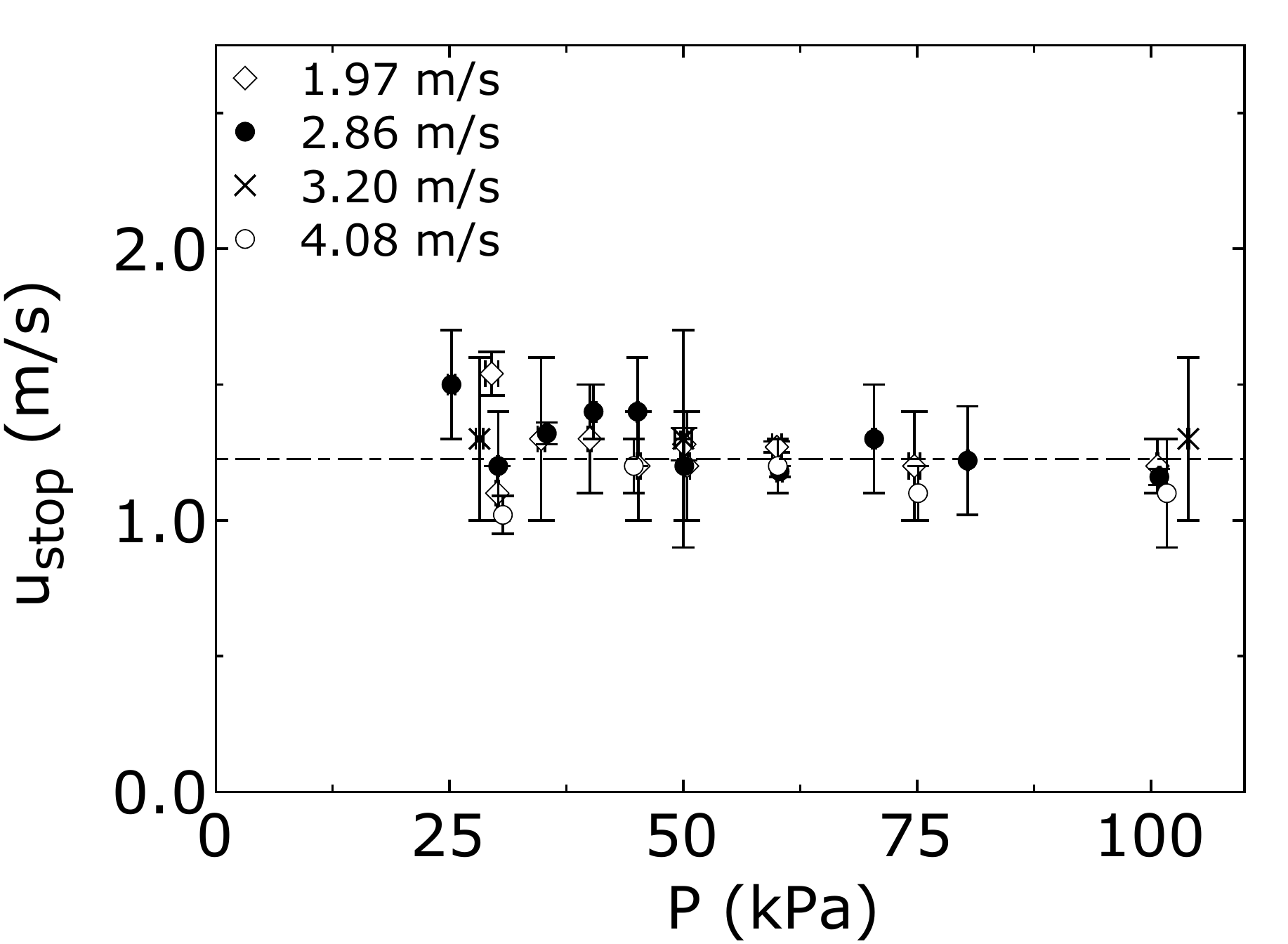} 
\end{center}
\caption{Rim velocity when air entrainment ceases, $u_{stop}$, versus pressure at different impact velocities. $u_{stop}$ is independent of both pressure and impact velocity.} 
\label{vstop}
\end{figure}

In order to understand what sets the time $t_{stop}$, we looked at other relevant scales in the problem.  As the drop expands, it  continually decelerates.  We measured the velocity of the lamella/sheet interface at the time $t_{stop}$ to obtain $u_{stop}$.  This is the lamella velocity at which bubble entrainment ceases. Figure \ref{vstop} shows $u_{stop}$ versus $P$ for a range of impact velocities.  As seen in Figure \ref{vstop}, $u_{stop}$ is independent of both pressure and impact velocity.  At any pressure above $P_{sh}$, once a thin sheet is present, bubbles are entrained  until the spreading velocity drops below $u_{stop}$.  Even though ambient gas pressure, $P$, is crucial for the creation of the instability which leads to bubble entrainment (the thin sheet), it does not appear to play a role in setting $t_{stop}$.

Bubble entrainment occurs only in the high $\nu_L$ splashing regime.  Close examination of impacts with 1 cSt silicone oil show no indication of bubble entrainment.  However above 3 cSt, the threshold for entering the high $\nu_L$ regime, bubble entrainment is always seen.  Bubble entrainment is unique to the high $\nu_L$ splashing regime, and is intrinsically linked with the presence of the thin sheet.  

The bubble entrainment we observe is reminiscent of what occurs in forced wetting processes.  This connection has been suggested previously \cite{Rein,DriscollDFD09,Siggi}.  In forced wetting, a solid is typically plunged into a bath.  Once the plunging velocity is above a critical value, cusps develop in the interface, and air bubbles are entrained behind these cusps \cite{Blake}.  One explanation for these shapes is that the contact line aims to minimize the normal velocity relative to the encroaching liquid \cite{Benkreira}.  The bubbles entrained in viscous splashing seem at least superficially similar in that they appear only above a threshold velocity.  However, we note several differences.  The moving liquid front in the splashing case is steadily decelerating, while forced wetting is studied at a constant velocity.  Bubbles are entrained behind cusps in the forced wetting case, while in viscous splashing entrainment occurs due to local contact of the thin sheet with the substrate.  Furthermore, the spreading edge is different in the two cases --- there is no thin sheet in the forced wetting process.  Nevertheless, the similarities between bubble entrainment in splashing and forced wetting are compelling.

\section{Conclusion}
There are several distinct regimes in the splashing of a drop on a smooth, dry substrate.  Although low-viscosity  and high-viscosity splashing appear to be markedly different, they both exhibit two separate regimes of behavior at high and low impact velocity.    As parameter space continues to be explored, even more splashing regimes may be found.

We have explored the high-viscosity splashing regime.  Before any drops break-off, the spreading liquid must first eject a thin sheet.  The threshold pressures for sheet ejection, $P_{sh}$, and drop break-off, $P_{br}$, depend on the gas molecular weight and impact velocity.  Increasing the liquid viscosity causes a widening gap between $P_{br}$ and $P_{sh}$, until above $30$ cSt, a thin sheet is produced, but no droplet break-off is observed even at atmospheric pressure.  

Viscous splashing occurs at delayed times compared to splashing at low viscosity \cite{LeiPRE}.  As gas pressure is lowered, $t_{ejt}$ increases until no sheet is emitted below $P_{sh}$.  We also find that $t_{ejt}$ scales with impact velocity, liquid viscosity, and gas molecular weight.  

We have shown that bubble entrainment occurs at the interface between the spreading lamella and the thin sheet.  It begins when the thin sheet is formed, and continues until the spreading lamella decelerates below the velocity $u_{stop}$, which is independent of both pressure and impact velocity. Bubble entrainment is a feature of high viscosity drop impacts which we have not observed to occur in the low-viscosity regime.

As we have shown, viscous splashing exhibits a rich behavior.  At this time, the basic mechanism for creating a viscous splash remains unknown.  Ejection of a thin liquid sheet is the precursor to drop break-off --- any proposed theory for splashing in this regime would need to account for its formation and evolution.  The effect of surface roughness in the high viscosity regime, as well as the size scales of the thin sheet compared to the lamella could offer clues as to what physical effects are contributing to the sheet ejection. 

\begin{acknowledgments} 
We are particularly grateful to Lei Xu, Wendy Zhang, and Robert Schroll.  We thank Nathan Keim, Joseph Paulsen, Arianna Strandburg-Peshkin, Justin Burton, and John Phillips for important advice, discussions, and help with experimental setup.  This work was supported by NSF grant No. DMR-0652269.  Use of facilities of the University of Chicago NSF-MRSEC and the Keck Initiative for Ultrafast Imaging are gratefully acknowledged. 
\end{acknowledgments}

\bibliography{splashingbib}

\end{document}